\documentclass[a4paper]{article}

\usepackage{INTERSPEECH2021}
\usepackage[dvipsnames]{xcolor}

\title{Voice of Your Brain: Cognitive Representations of Imagined Speech,\\
Overt Speech, and Speech Perception Based on EEG}
\name{Seo-Hyun Lee$^{1\dagger}$, Young-Eun Lee$^{1\dagger}$, Seong-Whan Lee$^{1,2*}$ \thanks{\\$^\dagger$Equal contribution\\ $^*$Corresponding author: Seong-Whan Lee}}
\address{
  $^1$The Department of Brain and Cognitive Engineering, Korea University\\
  $^2$The Department of Artificial Intelligence, Korea University}
\email{\{seohyunlee, ye\_lee, sw.lee\}@korea.ac.kr}

\begin{document}

\maketitle
\begin{abstract}
Every people has their own voice, likewise, brain signals display distinct neural representations for each individual. Although recent studies have revealed the robustness of speech-related paradigms for efficient brain-computer interface, the distinction on their cognitive representations with practical usability still remains to be discovered. Herein, we investigate the distinct brain patterns from electroencephalography (EEG) during imagined speech, overt speech, and speech perception in terms of subject variations with its practical use of speaker identification from single channel EEG. We performed classification of nine subjects using deep neural network that captures temporal-spectral-spatial features from EEG of imagined speech, overt speech, and speech perception. Furthermore, we demonstrated the underlying neural features of individual subjects while performing imagined speech by comparing the functional connectivity and the EEG envelope features. Our results demonstrate the possibility of subject identification from single channel EEG of imagined speech and overt speech. Also, the comparison of the three speech-related paradigms will provide valuable information for the practical use of speech-related brain signals in the further studies.

\end{abstract}

\section{Introduction}




Decoding speech from human brain signals have recently shown a significant potential of its practical use \cite{huth2016natural,schoffelen2017frequency}. Among the various speech-related paradigms (i.e., overt speech, speech perception),imagined speech is showing a significant potential in the field of brain-computer interface (BCI) communication \cite{wolpaw2002brain, lee2020neural, lee2018high}. Imagined speech is a concept of internal pronunciation of speech without an audible output nor the corresponding movement of the articulators \cite{schultz2017biosignal}. Although recent studies have shown the relevance and potential of imagined speech decoding \cite{lee2020neural, nguyen2017inferring}, the underlying neural characteristics and their practical usage are still remaining to be explored. However, previous studies have revealed the relevance of imagined speech with other speech-related concepts such as overt speech, or speech perception \cite{lee2019eeg, anumanchipalli2019speech}. Therefore, current studies on imagined speech are focusing on the common and contrast neural representations of speech-related paradigms, for a more robust and practical BCI.

The practical aspect of BCI technology is recently gaining attention with the current advances in BCI implying the potential of decoding human intention from brain signals \cite{yoon2020audio}. Although recognizing speaker from voice recordings is available with current speech technology \cite{jahangir2020text}, speaker identification from brain signals is emerging as a relevant issue due to the private and secure nature of brain signals \cite{dash2019spatial, moctezuma2020multi}. Several studies have shown the potential of subject identification from EEG of resting state, speech perception, or imagined speech conditions, however, have shown an inferior performance with multichannel EEG \cite{brigham2010subject}. For a practical use of brain signals, an improvement on the decoding performance as well as reducing the number of channels is required. Practical use of BCI systems has recently developed with the improvement of decoding performance using deep neural networks \cite{dash2019spatial}. Also, several attempts were made to decode human intention using practical EEG or single-channel EEG \cite{fiedler2017single, kwak2019error, lee2020real}. Speaker identification, therefore, carries issues to be more practical and secure by using single-channel \cite{fiedler2017single} or single trial \cite{dash2019spatial}.

Several deep learning techniques have been published to decode EEG brain signals, which are architectural designs that considers the characteristics of brain signal characteristics \cite{schirrmeister2017deep, lawhern2018eegnet}. It was often used to decode human intention using motor imagery or event-related potential, and have shown superior performance than the conventional machine learning methods such as linear discriminant analysis and support vector machine \cite{suk2012novel,kwon2019subject, gao2020classification}. Recently, there are several attempts to find optimal features of EEG by deep neural networks based on the three main features of EEG, temporal, spectral, and spatial features \cite{waytowich2018compact, bhatti2019soft}. In addition, EEG-based speaker identification studies also have actively applied machine learning or  deep learning techniques \cite{moctezuma2019subjects, dash2019spatial}. Deep learning may be effective in capturing prominent features from brain signals to verify individual characteristics.

In this paper, we investigated the EEG features of imagined speech in comparison with overt speech, speech perception, and resting state. Our main contributions were as follows:
(1) Three different speech related conditions (imagined speech, overt speech, and speech perception) and resting state EEG were compared in terms of speaker identification using temporal-spectral-spatial deep neural network.
(2) Our results demonstrate the brain signals of imagined speech and overt speech being individually characteristic, especially on single channel, implying the possibility of identifying individual speakers in practical situations. 
(3) We analyzed the functional connectivity and EEG envelope of each subject to verify the identification performance.



\section{Method}

\subsection{Data description}
The experimental protocol followed the previous works \cite{lee2020neural,lee2019eeg}. Nine subjects (three males; age 25.00 ± 2.96) participated in the study. The study was approved by the Korea University Institutional Review Board [KUIRB-2019-0143-01] and was conducted in accordance with the Declaration of Helsinki. Informed consent was obtained from all subjects.  

\begin{figure}[t]
\centering
    \includegraphics[width=\columnwidth]{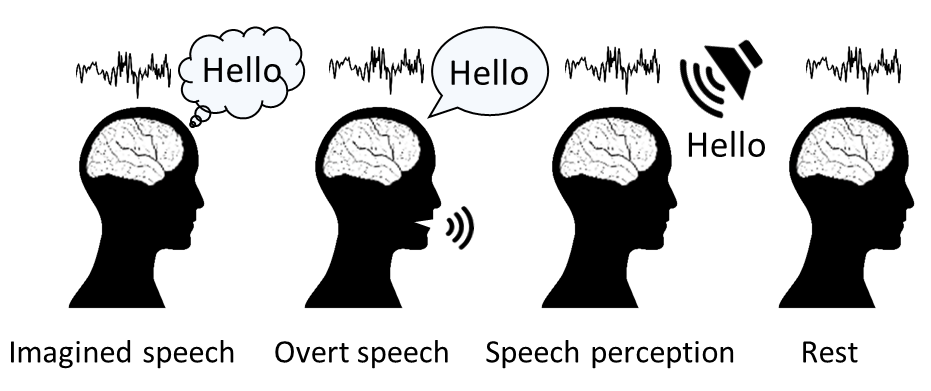}
    \caption{Four conditions (imagined speech, overt speech, speech perception, and resting state) analyzed in this study.}
    \label{fig1}
\end{figure}

EEG signals were acquired in four conditions, imagined speech, overt speech, speech perception, and resting state (Fig. \ref{fig1}). After recording two seconds of resting state, speech audio of each word/phrase was provided for another two seconds, followed by four consecutive trials of imagined speech or overt speech \cite{lee2020neural, lee2019towards}. Among the experiments that repeated the imagined or overt speech four times in a block, we used only the first trial from each block to match the number of trials with four experimental conditions. Each subject carried out twenty-five trials in random order for every twelve words, with a total of 300 trials for each condition.

\subsection{Data analysis}

\subsubsection{EEG preprocessing}
The EEG signals were down-sampled into 250 Hz and segmented into 2 s epochs from the beginning of each trial. The preprocessing of EEG signals were conducted with the fifth order Butterworth filter in the high-gamma band of 30--120 Hz and baseline was corrected by subtracting the average value of 500 ms before the trial onset. All the data processing procedures were conducted in Python and Matlab using OpenBMI toolbox \cite{leeMH2019eeg} and BBCI toolbox \cite{krepki2007berlin}.

\subsubsection{Classification framework}
\begin{figure}[t]
\centering
    \includegraphics[width=\columnwidth]{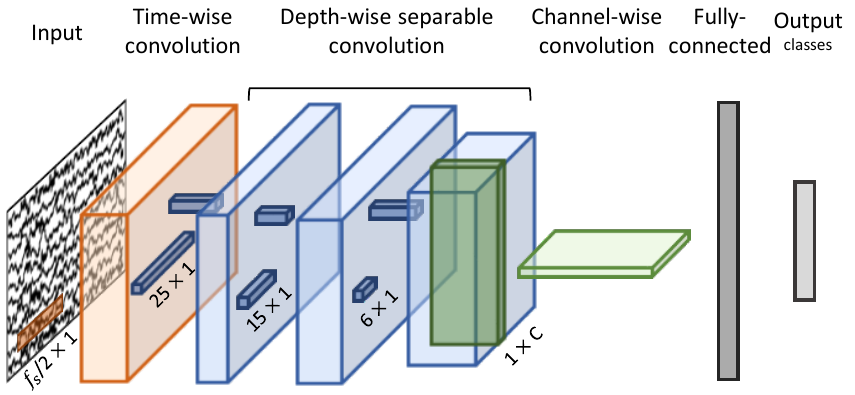}
    \caption{Classification framework based on temporal-spectral-spatial CNN.}
    \label{fig2}
\end{figure}

The proposed classification framework consists of convolution layers and separable convolution layers to extract the temporal-spectral-spatial information, as shown in Fig. \ref{fig2}. Given the input as raw signals (C $\times$ T), classification output is set to 9 speakers. The size of the kernel in the first layer is set in relation to the sampling frequency of the data to perform a temporal convolution that mimics band-pass filters \cite{waytowich2018compact}. Since support vector machine (SVM) classifier was reported to be robust in decoding imagined speech \cite{lee2020neural,nguyen2017inferring}, we used the squared hinge loss for the training which functions similar with the margin of SVM.
The evaluation was conducted with 5-fold cross-validation and training in 1000 epochs in each condition. The chance level of this experiment is 11.11\% as the number of samples of each subject were the same for imagined speech, overt speech, speech perception, and resting state conditions.

\subsubsection{Speaker identification}
Speaker identification was performed in each of the four conditions to distinguish speakers only with brain signals. Classification of each subject was performed with deep neural network described above which was designed to extract the individual EEG features of each subject. We compared the speaker identification performance of imagined speech, overt speech, speech perception, and resting state EEG in all channel and single channel environments. For the optimal single channel selection, we tested each channel located in the Broca's and Wernicke's area which are known to be highly related in the speech processing \cite{lee2020neural}. 
Furthermore, to verify the results, we tested our framework using speech perception and resting state EEG data collected in different sessions. 

\subsubsection{Functional connectivity}
The variations on the brain connectivity among the subjects were identified for understanding the cognitive representations of each individual when performing speech. We analyzed the phase locking value (PLV), which is a measure of functional connectivity that represents synchronization of electrical brain activities between two channels \cite{jann2009bold}. PLV was computed in imagined speech and resting state conditions using the channels located in the Broca and Wernicke’s areas (AF3, F3, F5, FC3, FC5, T7, C5, TP7, CP5, and P5). Specifically, we focused on the alterations in the brain connectivity by comparing the significant combinations of channels between imagined speech and resting state conditions.

\begin{figure*}[ht]
\centering
    \includegraphics[width=\textwidth]{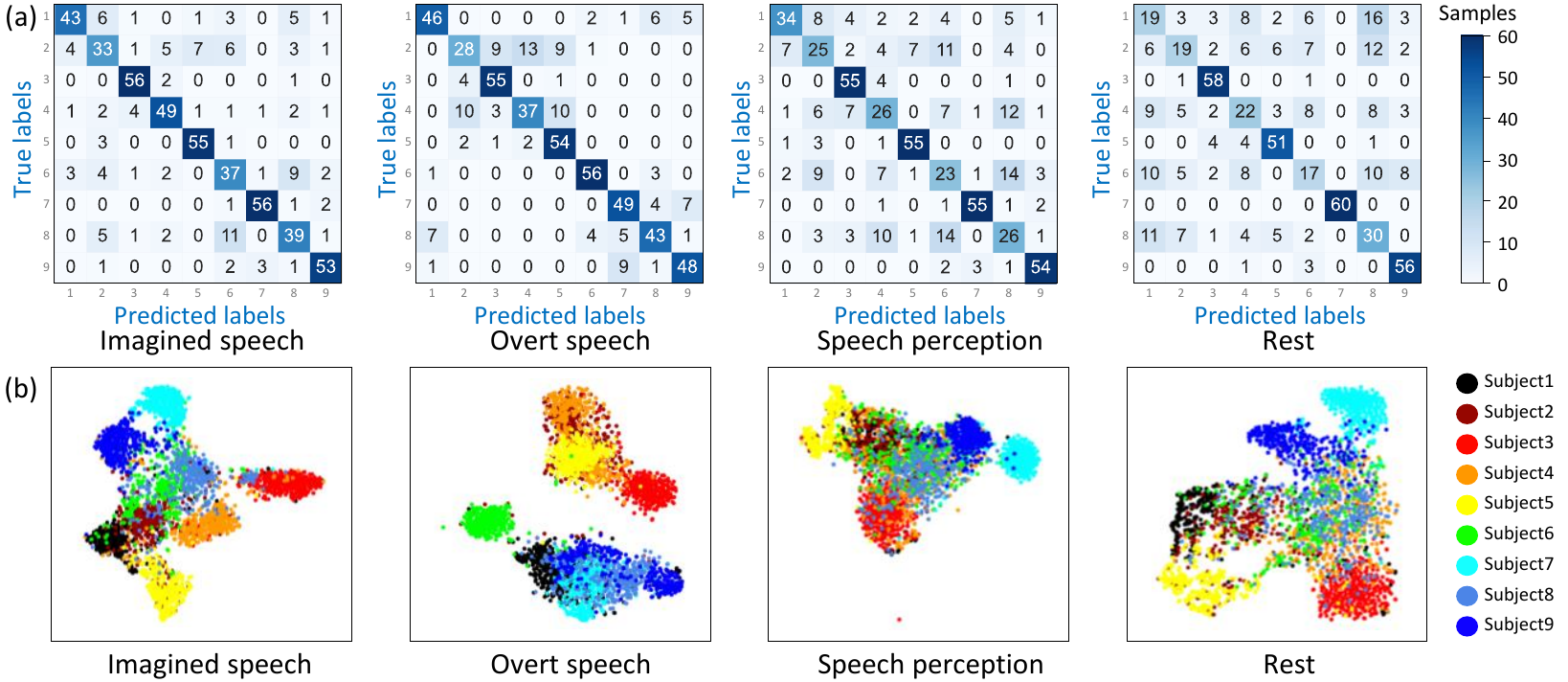}
    \caption{Classification results of single-channel EEG using temporal-spectral-spatial framework. (a) Normalized confusion matrix of imagined speech, overt speech, speech perception and rest. (b) T-SNE plot of dataset for each condition.}
    \label{fig3}
\end{figure*}

\begin{figure}[ht]
\centering
    \includegraphics[width=\columnwidth]{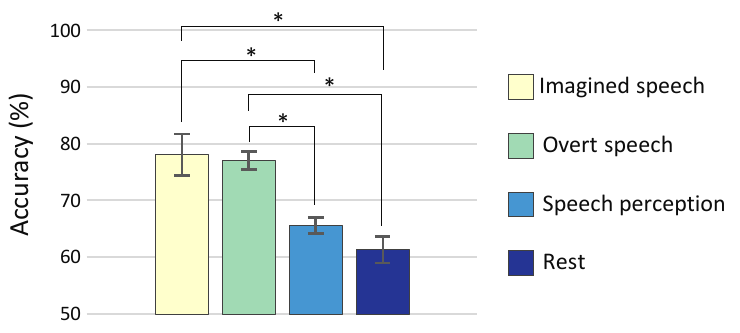}
    \caption{Classification performance of with or without convolution extracting spectral features of single channel and all channels.}
    \label{fig4}
\end{figure}

\subsubsection{EEG envelope}
The EEG envelope of imagined speech was analyzed to identify the presence of internal speech for each individual speaker. The envelope responds to the power changes corresponding to the phonemic transition, which is known to be the prominent representations of speech utterances reflected in the cortex \cite{horton2014envelope}. Temporal changes in the EEG envelope are known to reflect the cortical oscillations that corresponds to the cognitive process of loudness, pitch and timbre \cite{horton2014envelope}. The root mean square(rms) envelope was computed by each trials using 30-tap Hilbert filter and baseline corrected by 500 ms before the onset. We plotted the upper and lower value of the grand-averaged envelope of imagine speech and analyzed the distinction among subjects. 

\subsection{Statistical analysis}
The results were verified with statistical analysis. We performed Kruskal-Wallis non-parametric one-way analysis of variance (ANOVA) to compare the subject identification performance of imagined speech, overt speech, speech perception, and resting state conditions. Post-hoc analysis was performed with non-parametric permutation based t-test. Kruskal-Wallis test was also performed on the classification performance using single channel EEG to estimate the significance of the selected channel. Furthermore, a paired \textit{t}-test was performed to compare the PLV of imagery versus resting state EEG to find out the significant connectivity alterations in the Broca's and Wernicke's area while performing the imagined speech.

\section{Results and Discussions}

\subsection{Comparison of imagined speech, overt speech, speech perception, and resting state}

We investigated the speech-related EEG signals in four conditions of imagined speech, overt speech, speech perception, and rest. Fig. \ref{fig3} (a) displays the classification performance of single-channel EEG in four different conditions. As shown in the figure, subject 3 and 5 were well discriminated in all four conditions. However, while subject 6 have shown distinct features in overt speech condition, had relatively inferior characteristics in other conditions. This may imply the intensity of cognitive features reflected in brain signals may vary among different modes of speech. Fig. \ref{fig3} (b) shows the t-SNE plot of single-channel dataset of four conditions for each subject. Imagined speech and overt speech show relatively distinguished scatters of each speaker in the two-dimensional space, whereas speech perception and resting state have displayed ambiguous boundaries between speakers.

As shown in Fig. \ref{fig4}, imagined speech and overt speech showed relatively superior performance as 78.07\% and 77.07\% ($p=0.001$, $\chi^2=15.69$). Interestingly, the classification performance of imagined speech and overt speech did not differ significantly between each other ($p=0.56$, $t=0.63$), while overt speech was expected to show superior performance. Previous studies have shown relatively robust decoding performance under overt speech conditions \cite{lee2019eeg}, however, we could infer that the imagined speech also carries relevant information about the characteristics of each speaker. Another notable finding was that the identification performance of speech perception was significantly lower than that of imagined speech ($p=0.004$, $t=-5.86$) and overt speech ($p<0.001$, $t=-18.34$). Here, we could infer the possibility that active thinking or the utterance of speech rather than passive exposure to speech audio would be effective in discriminating speakers from brain signals. This may be a relevant finding since previous works on silent communication have highly conjugated the speech perception paradigm \cite{akbari2019towards}.

\begin{table}[]
\centering
\footnotesize
\renewcommand{\arraystretch}{1.1}
\caption{Single-channel classification performance of imagined speech in Broca's and Wernicke's area.}
\begin{tabular}[width=\columnwidth]{|c|c|c|c|c|c|}
\hline
\textbf{Channel}  & \textbf{F3} & \textbf{FC5} & \textbf{T7} & \textbf{CP5} & \textbf{AF3} \\ \hline
\textbf{Accuracy (\%)} & 77.04       & 76.19        & \textbf{78.07}       & 65.19        & 75.63        \\ \hline
\textbf{STD}      & 1.89        & 4.39         & \textbf{3.68}        & 1.18         & 1.96         \\ \hline
\textbf{Channel}  & \textbf{F5} & \textbf{FC3} & \textbf{C5} & \textbf{TP7} & \textbf{P5}  \\ \hline
\textbf{Accuracy (\%)} & 77.00       & 76.07        & 67.74       & 75.00        & 66.70        \\ \hline
\textbf{STD}      & 0.87        & 1.63         & 4.94        & 1.71         & 1.00         \\ \hline
\end{tabular}
\end{table}

\begin{figure}[th]
\centering
    \includegraphics[width=\columnwidth]{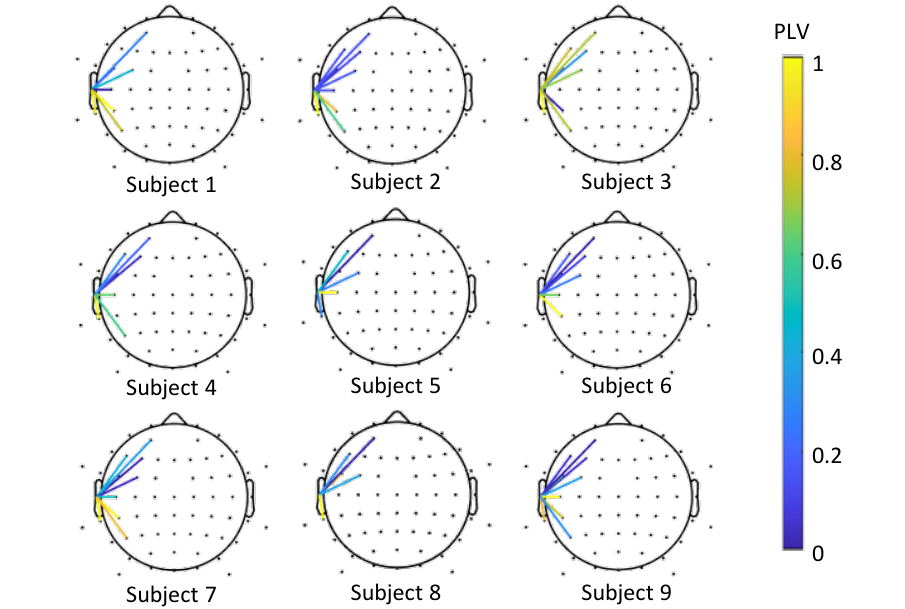}
    \caption{Phase locking value of imagined speech. Significant channels between T7 and each channel from Broca's and Wernicke's area were plotted by each subject.}
    \label{fig5}
\end{figure}

\subsection{Subject identification using single channel EEG}
Speaker identification from single-channel and multi-channel EEG signals was performed with the proposed architecture. When all channels were used, speakers were well identified regardless of speech-related paradigms (above 99.5 \%), which was in line with the results of the previous studies \cite{dash2019spatial, moctezuma2020multi}. Although brain signal decoding of classifying different words or sentences were limited in their actual use due to the inferior performance \cite{lee2020neural, nguyen2017inferring}, subject identification using EEG have shown the potential of its practical application owing to the robust performance \cite{dash2019spatial}. Here, we aimed to find an optimal EEG channel for subject identification. As shown in Table 1, T7 had shown significantly superior performance among the ten channels located in the speech-related areas ($p<0.001$, $\chi^2=32.51$). Since T7 is located near the ear, it represents a more practical channel, which displays the potential of subject identification using ear-EEG devices \cite{lee2020real}. In the multisession analysis, speech perception (56.26 ± 1.87 \%) had shown a superior performance than the resting state (53.87 ± 3.49 \%), however, further analysis is required since our dataset was limited in single session in the case of imagined speech and overt speech conditions. Also, session-to-session studies of BCI may further contribute in solving the multisession problems \cite{leeMH2019eeg}.


\subsection{Cognitive representations of speech}
Functional connectivity of each speaker during imagined speech was analyzed in Broca's and Wernicke's area, which are the cortical regions related to speech processing. We hypothesized that the distinguished brain connectivity while performing speech may be reflected as distinct features of each speaker. Fig. \ref{fig5} displays the significant PLV of imagined speech compared to that of the resting state. As shown in the figure, the functional connectivity of individual speakers have shown disparity among the subjects. Specifically, subject 3 had shown an increased connectivity in the Broca's area while most subjects have displayed decreased connectivity in the Broca's area. This can be interpreted along with Fig. \ref{fig3}, which exhibits discriminated features of subject 3 among other subjects. In addition, subject 4 who had relatively inferior true positive rate found in the Fig. \ref{fig3} (a) had the least channels that had shown significance between PLV of imagined speech and resting state. Meanwhile, an increased  brain connectivity was found in the Wernicke's area for the majority of subjects. Our results can be interpreted with the previous studies which have reported the decreased brain connectivity in the relevant areas during mental tasks \cite{jann2009bold, nguyen2017inferring}. This implies that the decreased PLV in the Broca's area reflects the active thought during imagined speech \cite{jann2009bold}.

\begin{figure}[t]
\centering
    \includegraphics[width=\columnwidth]{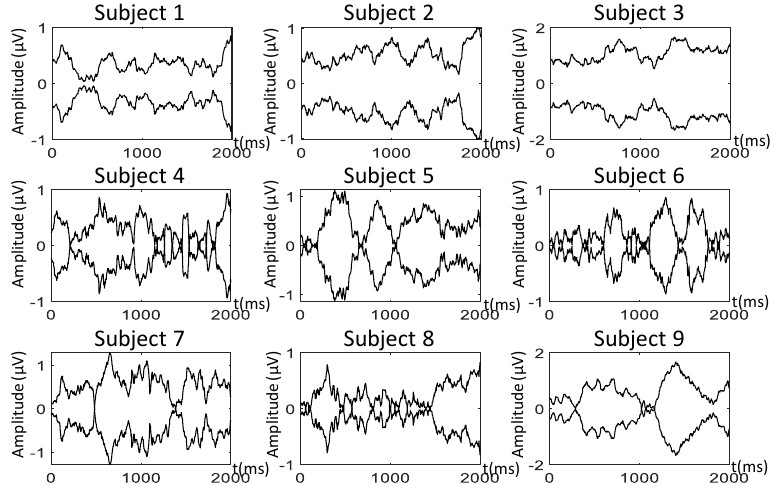}
    \caption{Grand averaged envelope of imagined speech EEG for each subject.}
    \label{fig6}
\end{figure}

Fig. \ref{fig6} displays the grand-averaged envelop of imagined speech for each subject.  
The EEG envelope of each subject showed the peaks around 500-1500 ms, while subject 3 exhibited a radiated feature pattern. Given the subject identification results, subject 3 was shown to be most distinctly classified in the t-SNE plot. We interpreted this results as subject 3 may have different temporal occurations of imagined speech since speech features of each speaker are also known to vary among each other \cite{horton2014envelope, lee2019eeg}.
The EEG envelopes can reflect distinct speech characteristics of each subject in brain signals, therefore may also be interpreted as voice of brain \cite{horton2014envelope, suk2011subject}.

\section{Conclusions}
We investigated brain signals of imagined speech, overt speech, and speech perception by classifying speakers using deep neural networks and demonstrated the individual neural representations by analyzing the brain connectivity and EEG envelope of inner voice. We analyzed the dataset of four speech-related conditions for the same subjects in this study. Speaker identification was performed from EEG signals of four conditions using the proposed CNN architecture that extracts temporal-spectral-spatial features. The distribution of samples going through the CNN was highly distinguished in the imagined speech and overt speech conditions. Imagined speech and overt speech were shown to incorporate relatively strong individual characteristics compared to the brain signals of speech perception or resting state. Furthermore, the connectivity and envelope of individual speakers during imagined speech implied that the speakers had individual characteristics reflected in their brain signals. Our results emphasize the potential of subject identification using single channel EEG of imagined speech. Further investigations on the distinct neural properties of individual speakers may further be interpreted as every speakers having their own inner voice during imagined speech.

\section{Acknowledgements}

This work is supported by the Institute for Information and Communications Technology Planning and Evaluation (IITP) grant funded by the Korea Government (MSIT) (Development of BCI-based Brain and Cognitive Computing Technology for Recognizing User’s Intentions Using Deep Learning) under grant 2017-0-00451.

\bibliographystyle{IEEEtran}

\bibliography{mybib}

\end{document}